\documentclass[preprint]{elsarticle}

\usepackage{amsmath}
\usepackage{epsfig}

\begin{document}

\title{Resampling and requantization of band-limited Gaussian stochastic signals
with flat power spectrum}

\author[ml]{Marco Lanucara}
\ead{marco.lanucara@esa.int}

\author[rb]{Riccardo Borghi}
\ead{borghi@uniroma3.it}

\cortext[cor1]{Corresponding author}

\address[ml]{European Space Operation Centre, European Space Agency, 
D-64293 Darmstadt, Germany}

\address[rb]{Dipartimento di Elettronica Applicata, Universit\`a degli Studi ``Roma Tre'', I-00146 Rome, Italy}

\begin{abstract}
A theoretical analysis, aimed at characterizing the degradation induced 
by the resampling and requantization processes applied to band-limited 
Gaussian signals with flat power spectrum, available through their digitized samples,
is presented.
The analysis provides an efficient algorithm for computing
the complete {joint} bivariate discrete probability distribution associated to the 
true quantized version of the Gaussian signal and to the quantity estimated
after resampling and requantization of the input digitized sequence.
The use of Fourier transform techniques allows deriving {approximate} analytical expressions for the quantities of interest,
as well as implementing their efficient computation. Numerical experiments are found to be in good agreement
with the theoretical results, and confirm the validity of the whole approach.
\end{abstract}

\begin{keyword}
quantization, interpolation, statistical signal processing 
\end{keyword}

\maketitle

\section{INTRODUCTION}
\label{intro}

Modern signal processing consists of algorithms applied to sequences of numbers, obtained 
by analogue to digital (A/D) conversion of analogue signals. 
The A/D conversion implies sampling in time domain and amplitude quantization, the second 
step being mandatory due to the finite length of the registers used for storing the samples 
amplitude in the processing machine. If the effect of quantization is disregarded, the exact 
reconstruction of the analogue signal from its samples is guaranteed by the sampling theorem,
under the assumption that the signal itself is band-limited.
Conversely, when quantization is applied, the exact reconstruction of the signal from the quantized samples is 
no longer possible.

An important signal processing task is the rate conversion applied to a sequence of numbers representing a digitized signal.
This task consists in obtaining samples of a signal taken at a certain rate, {say} $1/T_2$, based on the samples of the 
same signal available at a different rate, say $1/T_1$. This problem was extensively studied in the past years, for both rational and irrational values of the ratio $T_2/T_1$, assumed to be either larger (interpolation problem) or smaller (decimation problem) than unity~\cite{schafer,crochiere,crochiere2,ramstad}. The above cited papers
derive powerful techniques ensuring that the rate conversion is performed without degradation in all treated cases,
under the assumption that the signals are not quantized. 

In a non ideal condition, the input sequence available at rate $1/T_1$ and the output sequence obtained at rate $1/T_2$ 
as result of the rate conversion process are both quantized, in general (but not necessarily) according to the same quantization scheme. 
Requantization associated to rate conversion is applied in different contexts, like for example to
signals received from radio sources in many applications of radio astronomy~\cite{carlsonPASP-99,iguchiPASJ-05}, or to 
coded video data in image processing~\cite{japanese}.
In such cases it is of interest to establish theoretical bounds for the degradation 
occurring due to the quantization process, affecting both the input and the output sequences of numbers. 

The inclusion of quantization effects within the context of rate conversion was studied by the authors, in the specific case of extreme clipping, when only the sign of the analogue signal is recorded, i.e. when only one bit of information is associated to the amplitude of each sample~\cite{rockyI3ESP-07}. Under this hypothesis, and assuming that the input 
analogue signal is a realization $x(t)$ of a band-limited Gaussian process $X$ with flat power spectral density within the supporting bandwidth, results in closed form could be obtained about the degradation effect, in that context identified with the probability of error between the quantized version of $x(t)$ at any instant of time estimated through the available quantized samples, and the true quantized value of $x(t)$ (the ``target").

The present paper is devoted to extending the results of Ref.~\cite{rockyI3ESP-07} to the case of arbitrary quantization scheme, including multiple output levels, with the unique constraints of antisymmetry of the non-linear quantization function. The above mentioned probability of error, which was the metric used for quantifying the degradation effect in the binary case, is replaced by a complete bivariate discrete probability distribution or, in the case of large number of outputs, by the  cross-correlation coefficient between the estimated quantized value of $x(t)$ and the target.

\section{PRELIMINARIES}
\label{preliminaries}

Let $X$ be a stochastic process with real realizations $x(t)$, 
which is stationary and ergodic, with zero mean value and Gaussian statistics. 
The process is supposed to be band limited (BL for short), 
with flat power spectrum within the supporting bandwidth $[-W,W]$. 
On denoting by $\sigma$ the standard deviation of the process, it is 
well known that\cite{davenport}
\begin{equation}
\langle x(t_{1})\,x(t_{2}) \rangle = 
\sigma^{2}\,\textrm{sinc}\left(\displaystyle\frac{t_{2}-t_{1}}{T}\right),
\label{joint2D.3}
\end{equation}
where $T=1/2W$ is the inverse of the Nyquist frequency, the sinc 
function is defined by $\textrm{sinc}(\xi)=\sin(\pi \xi)/(\pi \xi)$,
and the symbol $\langle \cdot \rangle$ represents the expected value of its argument. 
Samples of the signal $x(t)$ are taken at known instants $kT$, so that 
$x_k = x(kT)$ denotes the $k$th sample. It is known that the signal 
$x(t)$ can be expanded (in a mean-square sense) as~\cite{davenport}
\begin{equation}
x(t)=\displaystyle\sum_{k=-\infty}^{+\infty}\,x_{k}\,\textrm{sinc}\left(\displaystyle\frac{t-kT}T\right).
\label{sampling}
\end{equation}

After the sampling, a quantization of the continuous value is 
performed, via a nonlinear function $f(x)$, so that
the final output, say the sequence $\{u_{m}\}$, can be expressed as 
follows:
\begin{equation}
    u_{m}= f(x_{m}), \,\,\,\,\,\,m=0,\pm 1,\pm 2, \dots
    \label{sampling.2.1}
\end{equation}
The function $f$ is assumed to be an antisymmetric, piecewise function
with an even number of outputs, i.e., none of the output levels equals zero. 
For $2M$ output levels we denote the (positive) discontinuity points by 
$0=a_1 <a_2 < \ldots <a_{M}<a_{M+1}=\infty$, as shown in Fig.~\ref{fig-levels}
where $f(x)$ is plotted for $x>0$. Note that the output levels $y_1<y_2<\ldots<y_M$ are 
also reported.

Of course, the case of quantization functions with odd number of levels
can be treated as well, by using the same methodology we are going to present.

We study the degradation associated to the reconstruction of the 
digitized version of the signal $x(t)$ at a time not belonging to the sampling grid $kT$,
based on the knowledge of its digitized samples $u_m$ given in Eq.~(\ref{sampling.2.1}). 
In view of the stationarity of the process $X$, such a problem consists in finding 
an estimate, say $\tilde u(\lambda)$, of the target value $f[x(\lambda\,T)]$, 
where $\lambda \in [0,1]$ is a dimensionless parameter.
To this aim, we first assume that a sinc interpolation is 
used to estimate $x(\lambda\,T)$, and then the quantization 
function $f$ is applied. The validity of this approach was demonstrated in Ref.~\cite{rockyI3ESP-07},
for any antisymmetric non linear function $f$.

Following the notations and results given in Ref.~\cite{rockyI3ESP-07},
the estimation of the digitized sample $\tilde u(\lambda)$ can be 
written as
\begin{equation}
	\begin{array}{l}
	\tilde u(\lambda)= f[w(\lambda)],
	\end{array}
    \label{sampling.2.3.1}
\end{equation}
where
\begin{equation}
    w(\lambda) = A_f\,\displaystyle\sum_{i=-\infty}^{+\infty}\,
    u_i\,\phi_i,
    \label{sampling.2.4.2.1}
\end{equation}
with  
\begin{equation}
    A_f=\displaystyle\frac{\langle x\,f(x)\rangle}{\langle f^2(x) 
    \rangle},
    \label{sampling.2.4.2.1.1}
\end{equation}
and
\begin{equation}
    \phi_i={\rm sinc}\,(\lambda-i).
    \label{sampling.2.4.2.1.2}
\end{equation}

In general, $\tilde u(\lambda)$ differs, even in a mean-square sense, from the target 
value $f[x(\lambda T)]$.
The effect of the degradation can be accounted for by determining the 
bivariate discrete probability distribution $p_{i,j}$,
equal to the probability that $\tilde u=y_j$ when the target is equal to
$y_i$, i.e.,
\begin{equation}
p_{i,j}={\rm Pr}\{f[x(\lambda T)]=y_i\,\,{\rm and}\,\,\tilde 
u(\lambda)=y_j\},
\label{bivariateDiscrete}
\end{equation}
for $i,j=\pm 1,\pm 2,\ldots,\pm M$.

The task of our analysis is therefore to
evaluate the discrete probability distribution of Eq.~(\ref{bivariateDiscrete}),
as a function of $\lambda$, for the most general case of $M \ge 1$.
The cross-correlation coefficient between the estimated and the target values will also be 
evaluated, from which the resampling and requantization-induced degradation could be easily inferred.

\section{THEORETICAL ANALYSIS}
\label{different}

The aim of the present section is to show the
theoretical basis of our approach for solving 
the problem stated in the previous section.
We retrieve the bivariate probability distribution in Eq.~(\ref{bivariateDiscrete})
by first evaluating its mixed moments up to the order 
$2M-1$, which is sufficient for the distribution
to be fully reconstructed. The subsequent step concerns the evaluation of the mixed moments,
which is achieved by employing a powerful and efficient method, 
making use of Fourier transform (FT for short) techniques.

Consider the mixed moments, say $\mu_{n,m}$, defined as 
\begin{equation}
\mu_{n,m}=\langle f(x)^n\,f(w)^m\rangle,
\label{moments}
\end{equation}
with $n,m=0,\ldots,2M-1$.
Note that, due the antisymmetry of $f$, the moments vanish whenever 
$n$ and $m$ have different parity.
The bivariate probability distribution $p_{i,j}$
can be arranged as a $2M \times 2M$ matrix, say $\mathbf{P}$,
which is defined as
\begin{equation}
      \mathbf{P}=
      \left[\
      \begin{array}{cccccc}
	p_{-M,-M} & \ldots & p_{-M,-1} & p_{-M,1} & \ldots & p_{-M,M}\\
 	&&&&&\\
	\ldots & \ldots & \ldots & \ldots & \ldots & \ldots\\
 	&&&&&\\
	p_{-1,-M} & \ldots & p_{-1,-1} & p_{-1,1} & \ldots & p_{-1,M}\\
 	&&&&&\\
	p_{1,-M} & \ldots & p_{1,-1} & p_{1,1} & \ldots & p_{1,M}\\
 	&&&&&\\
	\ldots & \ldots & \ldots & \ldots & \ldots & \ldots\\
 	&&&&&\\
	p_{M,-M} & \ldots & p_{M,-1} & p_{M,1} & \ldots & p_{M,M}\\
	\end{array}
      \right].
	\label{moments.2}
\end{equation}
By definition, the mixed moments are related to the bivariate distribution
according to the relation
\begin{equation}
\mu_{n,m}=\displaystyle\sum_{i,j}\,p_{i,j}\,y^n_i\,y^m_j,
\label{moments.2.1}
\end{equation}
which can be cast in a matrix form
\begin{equation}
      \boldsymbol{\mu}=\mathbf{Y}\,\mathbf{P}\mathbf{Y}^\dagger, 
	\label{moments.3}
\end{equation}
where the dagger denotes the transpose and the 2D matrices $\boldsymbol{\mu}$ and $\mathbf{Y}$ 
are defined by
\begin{equation}
      \boldsymbol{\mu}=
      \left[\
      \begin{array}{ccc}
	\mu_{0,0} & \ldots & \mu_{0,2M-1}\\
 	&&\\
	\ldots & \ldots & \ldots \\
 	&&\\
	\mu_{2M-1,0} & \ldots & \mu_{2M-1,2M-1}
	\end{array}
      \right],
	\label{moments.2.1.1}
\end{equation}
and
\begin{equation}
      \mathbf{Y}=
      \left[\
      \begin{array}{cccccc}
	1 & \ldots & 1 & 1& \ldots & 1\\
 	&&&&&\\
	y_{-M} & \ldots & y_{-1} & y_{1} & \ldots & y_{M}\\
 	&&&&&\\
	y^2_{-M} & \ldots & y^2_{-1} & y^2_{1} & \ldots & y^2_{M}\\
 	&&&&&\\
	y^3_{-M} & \ldots & y^3_{-1} & y^3_{1} & \ldots & y^3_{M}\\
 	&&&&&\\
	\ldots & \ldots & \ldots & \ldots & \ldots & \ldots\\
 	&&&&&\\
	y^{2M-1}_{-M} & \ldots & y^{2M-1}_{-1} & y^{2M-1}_{1} & \ldots & y^{2M-1}_{M}\\
	\end{array}
      \right],
	\label{moments.2.2}
\end{equation}
respectively.
Since $\mathbf{Y}$ is a Vandermonde matrix, and since all $y_i$'s are different, its inverse is always defined, so that 
the whole bivariate probability distribution $\mathbf{P}$
is trivially given by
\begin{equation}
      \mathbf{P}=\mathbf{Y}^{-1}\,\boldsymbol{\mu}(\mathbf{Y}^\dagger)^{-1}. 
	\label{moments.4}
\end{equation}
Concerning the correlation coefficient, this can also be derived from the knowledge
of the mixed moments defined in Eq.~(\ref{moments}) in the following way:
\begin{equation}
\rho=\displaystyle\frac{\mu_{1,1}}{\sqrt{\mu_{0,2}\,\mu_{2,0}}}.
\label{rho}
\end {equation}
The evaluation of the mixed moments $\mu_{n,m}$ pertinent to a typical 
$2M$-levels quantization function is not a trivial task.
Similarly to the approach used by Banta for evaluating autocorrelation functions
of quantized signals~\cite{banta}, we make use of a FT technique.
We start from the FT of the function $[f(x)]^n$, say $F_n(p)$, which is defined by
\begin{equation}
   [f(x)]^n=\displaystyle\int_{-\infty}^{+\infty}\,
   F_n(p)\,\exp(2\pi\textrm{i}xp)\,\mathrm{d}p,
    \label{ep.2}
\end{equation}
\textbf{
where $F_n(p)$ denotes the Fourier transform of $[f(x)]^n$, i.e.,
\begin{equation}
   F_n(p)=\mathcal{F}\{[f(x)]^n\}=\displaystyle\int_{-\infty}^{+\infty}\,
   [f(x)]^n\,
   \exp(-2\pi\textrm{i}xp)\,\mathrm{d}x,
    \label{ep.2.1}
\end{equation}
with $\mathcal{F}\{\cdot\}$ denoting the Fourier transform operator.
In Appendix \ref{app0} it is shown that
}
\begin{equation}
F_n(p)=
\left\{
\begin{array}{lr}
y^n_m\,\delta(p)+
\displaystyle\frac{1}{\pi p}\,\displaystyle\sum_{j=1}^{M-1}\,
\left(y^n_j-y^n_{j+1}\right)\,\sin(2\pi a_{j+1} p),& n\mbox{ even},\\
&\\
-\displaystyle\frac{{\rm i}y^n_1}{\pi p}+
\displaystyle\frac{{\rm i}}{\pi p}\,\displaystyle\sum_{j=1}^{M-1}\,
\left(y^n_j-y^n_{j+1}\right)\,\cos(2\pi a_{j+1} p),& n\mbox{ odd},
\end{array}
\right.
\label{FTquantization.1}
\end{equation}
where $\delta(p)$ denotes the Dirac distribution.

\textbf{
As far as the moment $\mu_{n,m}$ is concerned, 
on susbstituting from Eq. (\ref{ep.2}) into
Eq. (\ref{moments}) we have
\begin{equation}
    \mu_{n,m}= 
    \langle 
    \displaystyle\int_{-\infty}^{+\infty}\,
    \displaystyle\int_{-\infty}^{+\infty}\,
    F_n(p)\, F_m(p')\,
    \exp[\mathrm{i}2\pi x(\lambda T)p]\,
    \exp[\mathrm{i}2\pi w(\lambda)p']\,
    \mathrm{d}p\,\mathrm{d}p'
        \rangle ,
    \label{FTquantization.1.1}
\end{equation}
where the dependence on $\lambda$ has been made explicit.
Finally, on interchanging the integrals with the averages,
we obtain
}
\begin{equation}
    \mu_{n,m}= \displaystyle\int_{-\infty}^{+\infty}\,
    \displaystyle\int_{-\infty}^{+\infty}\,
    F_n(p)\, F_m(p')\,\langle 
    \exp\{2\pi\textrm{i}[x(\lambda T)p+w(\lambda)p']\}\rangle\,
    \mathrm{d}p\,\mathrm{d}p'.
    \label{ep.3}
\end{equation}
The quantity in the average can be written, starting from Eqs.~(\ref{sampling}) and~(\ref{sampling.2.4.2.1}), as
\begin{equation}
    \begin{array}{l}
    \langle 
    \exp\{2\pi\mathrm{i}[x(\lambda T)p+w(\lambda)p']\}\rangle
    =\langle 
    \exp\left\{2\pi\mathrm{i}\,\displaystyle\sum_{k} z_{k}\right\}\rangle,
    \end{array}
    \label{ep.4}
\end{equation}
where the zero-mean, \textbf{statistically independent}, random variables 
\begin{equation}
z_{k}=\phi_{k}\left[x_{k}\,p+A_{f}\,f(x_{k})\,p'\right],
\label{zetak}
\end{equation}
have been defined. 
Moreover, due to the above statistical independence, we have
\begin{equation}
\langle \exp({\rm i}2\pi Z)\rangle=
\displaystyle\prod_{k=-\infty}^{+\infty}\,
\langle \exp({\rm i}2\pi z_k)\rangle,
\label{charFunction.1}
\end{equation}
where 
\begin{equation}
Z=\displaystyle\sum_{k=-\infty}^{+\infty}\,z_k.
\label{charFunction.1.1}
\end{equation}
As we will see shortly, the average in the r.h.s. of Eq.~(\ref{charFunction.1}) 
can be calculated, for any $k$, for the considered class of quantization functions. 
However, the presence of the infinite product does not allow an exact closed form for the mixed moments
to be provided and makes their numerical estimation cumbersome. In order to overcome such difficulties, 
we are going to implement suitable approximations which will simplify the derivation of the moments.
 
\textbf{
First of all, it should be noted that the random variables $z_k$ are not normally distributed and, 
due to the prefactor $\phi_k$, are also not identically distributed, being 
$\sigma^2_{z_k} \to 0$ for $k\to\pm \infty$. The evaluation of the probability distribution function of infinite sums of 
not identically distributed indipendent random variables represents a task far from being trivial, as 
witnessed by past and current literatures\cite{petrov,lifshitsAP-97,crandall,schmulandAMM-03,rozovskyJMS-06,aurzadaJTP-07,borovkovJTP-08}, and is beyond the scope of the present work. In particular, it is clear that the use of the central limit theorem cannot be
invoked to find the probability density function of $Z$. To give a simple evidence of this, 
consider the case $\lambda=0$, for which $\phi_k=\delta_{0,k}$, and thus $Z=z_0$ that, as said above, 
is not normally distributed, unless $p'=0$. In the general case $\lambda\ne 0$, we use
the approach outlined in Ref. \cite{schmulandAMM-03}, so that the random variable $Z$ is first
written as the sum of two, statistically independent, random variables, say $Z_C$ and $Z_I$, defined as
\begin{equation}
    Z_{C}=\displaystyle\sum_{k\in {\cal N}} z_{k},
\label{charFunction.2.1a}
\end{equation}
and
\begin{equation}
    Z_{I}=\displaystyle\sum_{k\notin {\cal N}} z_{k},
       \label{charFunction.2.1b}
\end{equation}
respectively, with ${\cal N}$ being a suitable finite set of $N$ consecutive indices,
${\cal N}=\{i_1, i_2, \ldots,i_N\}$. In particular, we choose 
\begin{equation}
\mathcal{N}(h)=
\left\{
\begin{array}{lr}
\{-h+1,-h+2,\ldots,h-1,h\},&h>0\\
&\\
\emptyset,&h=0,
\end{array}
\right.
\label{setN}
\end{equation}
with $N=2h$ being the number of consecutive indices forming the set.\footnote{The explicit dependence of the set $\mathcal{N}$ on  the variable $h$ will not be shown in the subsequent formulas.}
In the above decomposition, the variable $Z_C$ retains the coefficients
$\phi_k$ for which $k$ is close to the interval of interest of $\lambda$,
while the variable $Z_I$ 
corresponds to the tails of the sum in Eq. (\ref{charFunction.1.1}), with
coefficients $\phi_k$ such that $|\phi_k(\lambda)|$ is slowly varying with respect to $k$, being
\begin{equation}
|\phi_k(\lambda)| \le 
\displaystyle\frac{\lambda}{|\lambda-k|},
\label{inequality}
\end{equation}
which, for large values of $|k|$, goes like $1/|k|$.
Following Ref. \cite{schmulandAMM-03}, we assume that $Z_I$
retains approximately a normal distribution, so that
}
\begin{equation}
\langle \exp(\mathrm{i}2\pi Z_I)\rangle
\simeq
\exp(-2\pi^2\sigma^2_{Z_I}),
\label{charFunction.3.0.0.0}
\end{equation}
where the variance $\sigma^2_{Z_I}=\langle Z^2_I \rangle$ is
\begin{equation}
\sigma_{Z_{I}}^{2}=
p^{2} P_{I} +  Q^2_{I}\,
(p^{\prime 2} + 2\,p\,p' ),
\label{charFunction.3.1a}
\end{equation}
with
\begin{equation}
P_{I} =\displaystyle\sum_{k \notin \mathcal{N}}\,\phi_{k}^{2} \langle x_{k}^{2}\rangle=
\sigma^{2}\,\left(1-\displaystyle\sum_{k \in \mathcal{N}}\,\phi_{k}^{2}\right),
\label{charFunction.3.1b}
\end{equation}
\begin{equation}
Q^2_{I}=A_{f}^{2}\,\displaystyle\sum_{k \notin \mathcal{N}}\,\phi_{k}^{2} \langle f^{2}(x_{k})\rangle=
A_{f}^{2}\,\langle f^2 \rangle\,\left(1-\displaystyle\sum_{k \in \mathcal{N}}\,\phi_{k}^{2}\right),
\label{charFunction.3.1c}
\end{equation}
where the fact that $\displaystyle\sum_{k}\,\phi_{k}^{2}=1$ has been used.
{In Sec. \ref{simulations}, in order to minimize the computational effort,
we have used the above decomposition with the minimum possible value of $h$, i.e., $h=1$ and 
$\mathcal{N}=\{0,1\}$. As we will see, such a choice provides quite satisfactory results.
}
As far as the $Z_C$ is concerned, 
{this variable is now defined as a sum of a \emph{finite} number of terms, which makes not
prohibitive the computation of the quantity $\langle\exp(\mathrm{i}2\pi Z_C)\rangle$, 
as required by Eq.~(\ref{charFunction.2}).
}
In particular, in Appendix~\ref{appendix.1} it is shown that
\begin{equation}
\begin{array}{l}
\langle\exp({\rm i}2\pi Z_C)\rangle=
\exp\left(-2\pi^2\sigma^2p^2\displaystyle\sum_{k\in{\cal N}}\,\phi_k^2\right)\\
\\
\times
\displaystyle\prod_{k\in {\cal N}}\,
\displaystyle\sum_j\,
\displaystyle\sum_s\,
\displaystyle\sum_q\,\displaystyle\frac 12\,
(-1)^{q}\,\exp({\rm i}2\pi\,s\,A_f\,y_j\,\phi_k\,p')\\
\\
\times\,{\rm erf}\left(\displaystyle\frac{a_{j+q}-{\rm i}\,2\pi\,s\,\phi_k\,\sigma^2\,p}{\sigma\sqrt 2}\right),
\end{array}
\label{charFunction.4}
\end{equation}
where ${\rm erf}(\cdot)$ denotes the error function~\cite{abramowitz},
$j=1,...,M$, $s\in\{-1,+1\}$, and $q\in\{0,1\}$.

\begin{equation}
    \langle \exp(2\pi\mathrm{i}\,Z)\rangle=
    \langle 
    \exp(2\pi\mathrm{i}\,Z_{C})\rangle
    \langle \exp(2\pi\mathrm{i}\,Z_{I})\rangle,
    \label{charFunction.2}
\end{equation}
where

Accordingly, on recalling Eq. (\ref{charFunction.3.0.0.0}), through Eqs.~(\ref{charFunction.3.1a})-(\ref{charFunction.3.1c})
we eventually found
\begin{equation}
\begin{array}{l}
\langle\exp({\rm i}2\pi Z)\rangle \approx
\exp\left[-2\pi^2(\sigma^2\,
p^2+ Q^2_I \,p^{\prime 2}+2 Q^2_I \,p\,p')\right]\\
\\
\times
\displaystyle\prod_{k\in {\cal N}}\,
\displaystyle\sum_j\,
\displaystyle\sum_s\,
\displaystyle\sum_q\,\displaystyle\frac 12\,
(-1)^{q}\,\exp({\rm i}2\pi\,s\,A_f\,y_j\,\phi_k\,p')\\
\\
\times\,{\rm erf}\left(\displaystyle\frac{a_{j+q}-{\rm i}\,2\pi\,s\,\phi_k\,\sigma^2\,p}{\sigma\sqrt 2}\right).
\end{array}
\label{charFunction.5}
\end{equation}

It is not difficult to show that, once Eqs.~(\ref{charFunction.5})
and~(\ref{FTquantization.1}) are substituted into Eq.~(\ref{ep.3}), 
the mixed moments take the following form (see Appendix~\ref{computational}):
\begin{equation}
    \begin{array}{lr}
    \mu_{n,m}=
    y_{M}^{n+m}+
    \displaystyle\sum_{j=1}^{M-1}\,y_{M}^{m}\,(y_{j}^{n}-y_{j+1}^{n})\,\mathcal{I}^{(e,1)}(\hat a_{j+1})\\
    \\+
    \displaystyle\sum_{j=1}^{M-1}\,y_{M}^{n}\,(y_{j}^{m}-y_{j+1}^{m})\,\mathcal{I}^{(e,2)}(\hat a'_{j+1})\\
    \\+
    \displaystyle\sum_{j=1}^{M-1}\,
    \displaystyle\sum_{j'=1}^{M-1}\,
    (y_{j}^{n}-y_{j+1}^{n})\,(y_{j'}^{m}-y_{j'+1}^{m})\,\mathcal{I}^{(e,3)}(\hat a_{j+1},\hat a'_{j'+1}),\\
    \end{array}
    \label{mixedMomentsEven}
\end{equation}
for even $n$ and $m$ and
\begin{equation}
    \begin{array}{lr}
    \mu_{n,m}=
    y_{1}^{n+m}\,\mathcal{I}^{(o)}(0,0)\\
    \\-
    \displaystyle\sum_{j=1}^{M-1}\,y_{1}^{m}\,(y_{j}^{n}-y_{j+1}^{n})\,\mathcal{I}^{(o)}(\hat a_{j+1},0)\\
    \\-
    \displaystyle\sum_{j=1}^{M-1}\,y_{1}^{n}\,(y_{j}^{m}-y_{j+1}^{m})\,\mathcal{I}^{(o)}(0,\hat a'_{j+1})\\
    \\+
    \displaystyle\sum_{j=1}^{M-1}\,
    \displaystyle\sum_{j'=1}^{M-1}\,
    (y_{j}^{n}-y_{j+1}^{n})\,(y_{j'}^{m}-y_{j'+1}^{m})\,\mathcal{I}^{(o)}(\hat a_{j+1},\hat a'_{j'+1}),\\
    \end{array}
    \label{mixedMomentsOdd}
\end{equation}
for odd $n$ and $m$, where the functions $\mathcal{I}^{(o)}(\cdot,\cdot)$,
 $\mathcal{I}^{(e,1)}(\cdot)$,  $\mathcal{I}^{(e,2)}(\cdot)$, and  $\mathcal{I}^{(e,3)}(\cdot,\cdot)$,
together with the symbols $\hat a_j$ and $\hat a'_{j+1}$ are defined in Appendix~\ref{computational}.

\section{THEORETICAL RESULTS}
\label{results}

The evaluation of the moments could be carried
out, through Eqs.~(\ref{mixedMomentsEven}) and~(\ref{mixedMomentsOdd}),
for arbitrary antisymmetric quantization functions of the form of Fig.~\ref{fig-levels}.
However, in the examples we are going to show, we
used the quantization schemes identified by Max
as the results of an optimization process
aimed at minimizing the distortion resulting from quantization~\cite{maxIREIT-60}.
For convenience, the geometries of the above quantization schemes are
reported in Tab.~\ref{table.1} for  values of $M$ up to 4.


Presenting the results related to the whole bivariate
probability distribution is a nontrivial task
due to the discrete and 2D character of the distribution
itself. We decided to present two examples of the $p_{i,j}$ distribution
for $M=4$ and for two fixed values of $\lambda$,
namely $\lambda=0.05$ and $\lambda=0.5$, which are reported
in Tabs.~\ref{table.2a} and~\ref{table.2b}, respectively. 

A meaningful parameter quantifying the
amount of degradation induced by the quantization and resampling
processes is the cross-correlation coefficient, defined in Eq.~(\ref{rho}),
whose behavior, as a function of $\lambda$, is plotted in 
Fig.~\ref{figCorrCoeff.1} for $M=1,\ldots,4$.
Note that, due to symmetry reasons, only the interval $[0,1/2]$ of $\lambda$
is shown.

As a general remark, it should be noted that, on increasing the number
of quantization levels, the correlation coefficient increases approaching 1
and displays a {\em plateau} whose extension approaches the whole $\lambda$
interval. Both behaviors are expected, and they account for the fact that,
for dense and non-clipping quantization schemes, $f(x) \to x$.


Although the methodology presented so far provides a complete solution 
to the problem under investigation, the involved computational effort
increases proportionally to $M^4$. As a matter of fact, its use 
for large values of $M$, i.e., for dense quantization schemes,
is made difficult by practical constraints related to the
computation time required for evaluating all involved integrals and
to the numerical stability of the final results.
In fact, a possible drawback occurs when the Vandermonde matrix in Eq.~(\ref{moments.2.2})
has to be inverted for large values of  $M$, to derive the full bivariate 
probability distribution. In this case, however, it is preferable to
deal with the problem in terms of the cross-correlation coefficient, which provides an adequate description 
of the degradation effect.

\section{COMPARISON WITH NUMERICAL SIMULATIONS}
\label{simulations}

We performed numerical simulations aimed at 
quantitatively verifying the theoretical results presented in Sec.~\ref{results}.

In Fig.~\ref{fig-scheme-sim} a schematic block diagram
explaining the methodology adopted for the simulations
is sketched.
A sequence of random numbers normally distributed
with unit variance and zero mean is generated,
representing the samples of a realization of
a Gaussian process $X$ taken at the sampling period which is
identified as 1 sec.
By construction, the process $X$ is BL between -1/2 and 1/2 Hz.
The samples $\{x_k\}$ are used, along two parallel signal paths,
to generate the values of $f[x(\lambda)]$ and $f[w(\lambda)]$
{according to the reconstruction formulas~(2) and~(5). Only a 
finite number of terms is used for reconstruction; the adopted 
selection of 200 terms is justified in Appendix~\ref{finiteSamples}.
The values  $f[x(\lambda)]$ and $f[w(\lambda)]$}
 are then used  to estimate the mixed moments $\mu_{n,m}$
and the discrete probability $p_{i,j}$, independently. 
In particular, $p_{i,j}$ has been evaluated by counting 
the events $f[x(\lambda)]=y_i$ and $f[w(\lambda)]=y_j$ for 
a large number of realizations (in the order on 10$^5$).
The mixed moments have also been estimated by averaging the product
$f^n(x)\,f^m(x)$ over the same number of realizations, 
in order to verify the theoretical predictions about the correlation
coefficient.

Tables~\ref{table.3a} and~\ref{table.3b} give 
the bivariate discrete probability distribution
estimated from numerical simulations corresponding
to the case $M=4$, for $\lambda=0.05$, and $\lambda=0.5$,
respectively. They have to be compared to
tables~\ref{table.2a} and~\ref{table.2b}, respectively.
As we can see, the agreement between the theoretical and
experimental probability distributions is very good.

As far as the correlation coefficient is concerned, 
Fig.~\ref{figSimulations.1} shows the 
extremely good agreement between the theoretical values of $\rho$ 
plotted in Fig.~\ref{figCorrCoeff.1} (solid curves) and the 
experimental results obtained by numerical simulations (circles).

Before concluding the present section, it is worth providing some 
details about the choice of the number of samples
used in the reconstruction formula of Eq.~(\ref{sampling.2.4.2.1}).
To this aim, Appendix~\ref{finiteSamples} contains a detailed 
analysis concerning the way the truncation of the series
in Eq.~(\ref{sampling.2.4.2.1}) affects the degradation of the reconstructed
signal. In particular it is confirmed, by suitable numerical experiments,
that a number of samples of about 200 is enough to validate the 
excellent agreement between theory and experiment previously displayed.

\section{AN APPLICATION TO DIGITAL SIGNAL PROCESSING}
\label{src}

In the present section we illustrate a practical application of our theoretical results. 
We review the problem of the sampling rate increase (interpolation), making reference to the classical theory by 
Schafer and Rabiner~\cite{schafer}. In particular we show how the degradation originated from the re-sampling and re-quantization processes can be accounted for by using the theoretical expressions presented 
in Sec.~\ref{different}.

We start from the top part of the block diagram of Fig.~\ref{fig-src.1}.
The signal $x(t)$, defined in Sec.~\ref{preliminaries}, 
is first sampled at the rate $1/T$. The obtained sequence $\{x_{k}\}$ is then 
interpolated and quantized, producing the output sequence $\{u_{m}\}$.
The sampling rate associated to the sequence $\{u_{m}\}$ is $1/T'$, 
where $T'/T = D/L$, with $D$ and $L$ being integer numbers greater than  1, with $D<L$.\footnote{We limit ourselves to 
the case of interpolation, for which $T' <T$.} 
The change of the sampling rate from $1/T$ to $1/T'$ is operated according to the 
prescriptions given in Ref.~\cite{schafer}. More precisely, after the first 
block the sampling rate is increased by the integer factor $L$,
by inserting a sequence of $L-1$ zero-valued 
samples between any two consecutive elements of the original 
sequence. The sequence so obtained is filtered through an ideal low-pass filter 
having a normalized cutoff frequency $\pi/L$ and gain $L$. The output of the filter, is 
decimated by selecting a sample every $D$ and eventually quantized by the function $f$.
Following Ref.~\cite{schafer}, it is possible to show that the relation between the input sequence
$\{x_k\}$ and the output sequence $\{u_m\}$ is given by
\begin{equation}
    u_{m}=f\left[\displaystyle\sum_{k}\,x_{k}\,\phi_{k}(\tilde\lambda_{m})\right]=f[x(mT')],
\label{src.2}
\end{equation}
where $\tilde\lambda_{m}=\displaystyle\frac DL m$.

In the bottom part of Fig.~\ref{fig-src.1}, the same processing scheme is adopted assuming 
that the signal $x(t)$ is only available through its digitized samples $\{f(x_k)\}$.
The input sequence $\{f(x_k)\}$ is suitably scaled by the normalization factor $A_f$,
which will be set to one, assuming the use of an ideal quantizer $f$~\cite{maxIREIT-60}.
The outcome of this processing is now represented by the sequence $\{\tilde u_m\}$, where
\begin{equation}
    \tilde u_{m}=f\left[\displaystyle\sum_{k}\,f(x_{k})\,\phi_{k}(\tilde\lambda_{m})\right].
\label{src.3}
\end{equation}
We note that Eq.~(\ref{src.3}) coincides with Eqs.~(\ref{sampling.2.3.1}) and~(\ref{sampling.2.4.2.1}),
so that we can apply our theoretical results to the present situation. 
Each element of the sequence $\{\tilde u_{m}\}$ is a degraded version of the
corresponding element of the sequence $\{u_{m}\}$ (the target).
Such a degradation is not stationary with respect to the ``time", represented by the index $m$. 
In fact, when $m D/L$ is an integer number there is indeed no degradation, whereas when the same 
quantity has fractional part equal to $1/2$, we know that the degradation is maximum (see Fig.~\ref{figCorrCoeff.1}).

The overall degradation between the two sequences can be quantitatively accounted for
by the $0$-delay temporal degree of coherence, say $\gamma$, which is defined by
\begin{equation}
    \gamma=\displaystyle\frac{\overline{u_m\,\tilde u_m}}
    {\sqrt{\overline{u^2_m}}\,
    \sqrt{\overline{\tilde u^2_m}}},
    \label{src.3.1}
\end{equation}
where the bar denotes the temporal (i.e., $m$) average.
The above definition can be used, provided that it is shown to be
independent of the particular realization $x(t)$.
After straightforward algebra, it is possible to prove the following 
theoretical expression for $\gamma$, in terms of the mixed moments ${\mu}_{n,m}$,
defined in Eq.~(\ref{moments}):
\begin{equation}
    \gamma=\displaystyle\frac{
    \displaystyle\frac 
    1L\,\displaystyle\sum_{i=0}^{L-1}\,\mu_{1,1}(\lambda_{i})
    }
    {
    \sqrt{\mu_{2,0}}\,
    \sqrt{\displaystyle\frac 
    1L\,\displaystyle\sum_{i=0}^{L-1}\,\mu_{0,2}(\lambda_{i})}\,
      },
    \label{src.4}
\end{equation}
where $\lambda_{i}=i/L$, $(i=0,\ldots,L-1)$, and the fact that $\mu_{2,0}$ does not depend 
on $\lambda$ has been made explicit. It should be noted that, for the treated interpolation problem,
only the value of $L$ is relevant for the degradation.

The quantity in Eq.~(\ref{src.3.1}) is easily measurable by implementing the block diagram of Fig.~\ref{fig-src.1},
whereas expression in Eq.~(\ref{src.4}) is purely theoretical, based on the expression of the moments obtained 
in Sec.~\ref{different}.
To provide the experimental verification of the identity between the two equations,
the various DSP blocks of Fig.~\ref{fig-src.1} have been implemented.
In particular, a FIR filter has been used to implement the filter block,
based on windowing the ideal impulse response corresponding to a 
rectangular transfer function by a Hamming window, similarly as we did in Ref.~\cite{rockyI3ESP-07}.
Values of $\gamma$ obtained for values of $L$  from 2 up to 30, and for
various values of $D$, have been experimentally evaluated and reported 
is Fig.~\ref{fig-src.2} 
\textbf{for $M=1,2,3,4$}. 
\textbf{Solid curves represent} the 
theoretical values provided by Eq.~(\ref{src.4}). 
The agreement is excellent. 

\section{CONCLUSIONS}
\label{conclusions}

The purpose of this paper was to extend the studies about the rate conversion 
applied to signal sequences, by taking into account the degradation effect 
associated to the quantization process.

In particular, we addressed the problem of computing the degradation induced by the 
resampling  and re-quantization of a BL stationary and ergodic signal with Gaussian 
statistics and flat power spectrum within the supporting bandwidth, available 
through its quantized samples. 

The analysis provides the algorithm for quantitatively characterizing the degradation effect
induced by the resampling and re-quantization processes in terms of the knowledge
of the complete bivariate discrete probability distribution associated to the target
and the estimated quantized signals, or in terms of the correlation coefficient
between the two quantities. The analysis makes use of FT representation of the quantization function and of its powers, to allow the application of linear analysis techniques.
Numerical experiments have also been implemented in order to validate the theoretical methodology
and results. The comparison showed an excellent agreement between theory and simulations.
Finally, we provided an example of application of our theoretical results to an important area
of Digital Signal Processing, the sampling rate conversion. 

The class of stochastic processes considered in the present paper represents a fundamental model with important applications in radio astronomy, where the received noise-like signal originated from radio sources can be modeled, after filtering, with good accuracy by the ideal process here analysed. In such applications coarse quantization (1 or 2 bits) is commonly applied, and  the obtained sequences are often subject to resampling and requantization. 
Of course, the results obtained in the present paper must be interpreted as a first step toward the extension of the methodology originally developed in Ref.~\cite{rockyI3ESP-07} to other important classes of stochastic signals. 
Within the same perspective another important topic, which will be the subject of forthcoming studies, concerns the development of an asymptotic analysis dealing with the case of dense and no-clipping  quantization schemes, aimed at deriving closed-form limit expressions describing the degradation effect.

\textbf{
Finally, we wish to suggest a possible, future extension of our work 
which also could be of interest for the radio-astronomy community.
In particular, we refer at the classical problem in very large baseline interferometry (VLBI), of different antennas 
observing the same object and producing digitized representations of the
received signal which are then cross-correlated to extract observables of
interest. The decorrelation induced by the quantization process has been
studied in various works in the past, for example in Ref. \cite{iguchiITC-02}.
We now think to the case, presently of practical interest, where the received
downlink signal is digitized at the different antennas according to different
sampling and quantization schemes. In such a case the data streams have to be
brought to a common quantization scheme and sampling rate prior to
cross-correlation. We believe that the additional de-correlation induced by
the re-sampling and re-quantization can be studied by use of the methods
exposed in our paper.
}

\section*{Acknowledgment}

We wish to thank Turi Maria Spinozzi for his invaluable help during all the phases of the preparation of 
the manuscript.

\appendix

\section{Derivation of Eq. (\ref{FTquantization.1})}
\label{app0}

\textbf{
We start from Eq. (\ref{ep.2.1}) which, by FT inversion, gives
\begin{equation}
   F_n(p)=
   \displaystyle\int_{-\infty}^{+\infty}\,
   [f(x)]^n\,\exp(-\mathrm{i}2\pi px)\,\mathrm{d}x,
    \label{app0.1}
\end{equation}
where, due to the piecewise character of the quantization function, 
\begin{equation}
[f(x)]^n=
\left\{
\begin{array}{lr}
y^n_M\,-
\displaystyle\sum_{j=1}^{M-1}\,
\left(y^n_{j+1}-y^n_j\right)\,
\mathrm{rect}\left(\displaystyle\frac x{2 a_{j+1}}\right),& n\,\mathrm{ even},\\
&\\
y^n_M\,\mathrm{sign}(x)-
\displaystyle\sum_{j=1}^{M-1}\,
\left(y^n_{j+1}-y^n_j\right)\,\mathrm{sign}(x)\,
\mathrm{rect}\left(\displaystyle\frac x{2 a_{j+1}}\right),& n\,\mathrm{odd},
\end{array}
\right.
\label{app0.2}
\end{equation}
where the function rect$(x)$ is defined as
\begin{equation}
\mathrm{rect}(x)=
\left\{
\begin{array}{lr}
1,& |x| \le 1/2,\\
&\\
0,& |x| > 1/2,
\end{array}
\right.
\label{app0.3}
\end{equation}
and sign$(x)$ denotes the signum function, i.e.,
\begin{equation}
\mathrm{sign}(x)=
\left\{
\begin{array}{lr}
1,& x >0,\\
&\\
-1,& x <0,\\
&\\
0,& x =0.
\end{array}
\right.
\label{app0.4}
\end{equation}
Furthermore, on taking into account that
\begin{equation}
\mathcal{F}\left\{\mathrm{rect}\left(\displaystyle\frac x{2a}\right)\right\}=
\displaystyle\frac{\sin(2\pi a p)}{\pi p},
\label{app0.5}
\end{equation}
and
\begin{equation}
\begin{array}{l}
\mathcal{F}\left\{
\mathrm{sign}(x)\,\mathrm{rect}\left(\displaystyle\frac x{2a}\right)
\right\}=\\
\\
=
\mathcal{F}\left\{
\mathrm{rect}\left(\displaystyle\frac {x-a/2}{a}\right)
\right\}+
\mathcal{F}\left\{
\mathrm{rect}\left(\displaystyle\frac {x+a/2}{a}\right)
\right\}=\\
\\
=-2\mathrm{i}\,
\sin(\pi a p)\,
\mathcal{F}\left\{
\mathrm{rect}\left(\displaystyle\frac {x}{a}\right)
\right\}=-\displaystyle\frac{2\mathrm{i}}{\pi p}\,\sin^2(\pi a p)=
\\
=
-\displaystyle\frac{\mathrm{i}}{\pi p}\,[1-\cos(2\pi a p)],
\end{array}
\label{app0.6}
\end{equation}
after some algebra Eqs. (\ref{app0.1})
and (\ref{app0.2}) lead to Eq. (\ref{FTquantization.1}).
}

\section{Derivation of Eq.~(\ref{charFunction.4})}
\label{appendix.1}

We start from
\begin{equation}
\langle\exp(2\pi\mathrm{i}\,Z_{C})\rangle=\displaystyle\prod_{k\in{\cal N}}\,\psi_k(p,p'),
\label{appendix.1.1}
\end{equation}
where
\begin{equation}
\begin{array}{l}
    \psi_k(p,p')=
    \langle \exp\{2\pi\mathrm{i}\,\phi_k\,[x_k\, p+A_f\,f(x_k)\,p']\}\rangle=\\
\\
=
\displaystyle\int_{-\infty}^{+\infty}\,p_x(x)\,\exp\{2\pi\mathrm{i}\,\phi_k\,[x\, p+A_f\,f(x)\,p']\}
{\rm d}x,
\end{array}
\label{appendix.1.2}
\end{equation}
where $p_x(x)$ is the pdf of the Gaussian process $X$.
Due to the piecewise character of the $f(x)$, Eq.~(\ref{appendix.1.2})
can be written as
\begin{equation}
\begin{array}{l}
    	\psi_k(p,p')=
	\displaystyle\sum_{j=1}^M\,
	\displaystyle\int_{a_j}^{a_{j+1}}\,p_x(x)\,\exp\{2\pi\mathrm{i}\phi_k\,[x\, p+A_f\,y_j\,p']\}\,{\rm d}x\\
\\
	+\displaystyle\sum_{j=1}^M\,
	\displaystyle\int_{-a_{j+1}}^{-a_{j}}\,p_x(x)\,\exp\{2\pi\mathrm{i}\phi_k\,[x\, p-A_f\,y_j\,p']\}\,{\rm d}x
.
\end{array}
\label{appendix.1.2.1}
\end{equation}
Furthermore, on changing the integration variable $x$ in $-x$ in the second integral, after trivial algebra 
we obtain
\begin{equation}
\begin{array}{l}
    	\psi_k(p,p')=
\\
=
	\displaystyle\sum_{j=1}^M\,
	\exp(2\pi\mathrm{i}\,\phi_k\,A_f\,y_j\,p')
	\displaystyle\int_{a_j}^{a_{j+1}}\,p_x(x)\,\exp(2\pi\mathrm{i}\phi_k\,x\,p)\,{\rm d}x+ {\rm c.c.},
\end{array}
\label{appendix.1.2.2}
\end{equation}
where c.c. stands for {\em complex conjugate}.
The last integral can be analytically expressed in terms of error function,
and precisely
\begin{equation}
\begin{array}{l}
	\displaystyle\int_{a_j}^{a_{j+1}}\,p_x(x)\,\exp(2\pi\mathrm{i}\phi_k\,x\,p)\,{\rm d}x=
\displaystyle\frac 12\,\exp(-2\pi^2\phi^2_k\sigma^2p^2)\\
\\
\times\left[{\rm erf}\left(\displaystyle\frac{a_{j+1}-2\mathrm{i}\pi\phi_k\sigma^2p}{\sqrt{2}\sigma}\right)-
{\rm erf}\left(\displaystyle\frac{a_{j}-2\mathrm{i}\pi\phi_k\sigma^2p}{\sqrt{2}\sigma}\right)\right].
\end{array}
\label{appendix.1.2.3}
\end{equation}
On substituting from Eq.~(\ref{appendix.1.2.3}) into Eq.~(\ref{appendix.1.2.2}), 
and on introducing two binary indices, say $q\in\{0,1\}$ and $s\in\{-1,1\}$, we have
\begin{equation}
\begin{array}{l}
    	\psi_k(p,p')=-\exp(-2\pi^2\phi^2_k\sigma^2p^2)\\
\\
	\times\displaystyle\sum_{j,q,s}\,\displaystyle\frac {(-1)^q}2\,\exp(2\pi\,\mathrm{i}\,s\,\phi_k\,A_f\,y_j\,p')\,
      {\rm erf}\left(\displaystyle\frac{a_{j+q}-2\mathrm{i}\,s\,\pi\phi_k\sigma^2p}{\sqrt{2}\sigma}\right).
\end{array}
\label{appendix.1.2.5}
\end{equation}
Finally, on substituting Eq.~(\ref{appendix.1.2.5}) into Eq.~(\ref{appendix.1.1}),
after trivial algebra Eq.~(\ref{charFunction.4}) naturally follows.

\section{Some computational remarks}
\label{computational}

First of all, we note that Eq.~(\ref{charFunction.5}) can be formally rewritten in the following way:
\begin{equation}
\begin{array}{l}
\langle\exp({\rm i}2\pi Z)\rangle=\,\displaystyle\frac 1{2^M}\,
\exp\left[-2\pi^2(\sigma^2\,
p^2+ Q^2_I \,p^{\prime 2}+2 Q^2_I \,p\,p')\right]\\
\\
\times
\displaystyle\sum_{{\bf j}}\,
\displaystyle\sum_{{\bf s}}\,
\displaystyle\sum_{{\bf q}}\,
\displaystyle\prod_{k\in {\cal N}}\,
(-1)^{qk}\,\exp({\rm i}2\pi\,s_k\,A_f\,y_{j_k}\,p')\\
\\
\times\,{\rm erf}\left(\displaystyle\frac{a_{j_k+q_k}-{\rm i}\,2\pi\,s_k\,\phi_k\,\sigma^2\,p}{\sigma\sqrt 2}\right),
\end{array}
\label{charFunction.6}
\end{equation}
where the vectorial indices ${\bf j}$, ${\bf s}$, and ${\bf q}$ are defined by
\begin{equation}
\begin{array}{l}
{\bf j}=[j_{i_1},j_{i_2},\ldots,j_{i_{N-1}},j_{i_{N}}],\\
\\
{\bf s}=[s_{i_1},s_{i_2},\ldots,s_{i_{N-1}},s_{i_{N}}],\\
\\
{\bf q}=[q_{i_1},q_{i_2},\ldots,q_{i_{N-1}},q_{i_{N}}],
\end{array}
\label{charFunction.7}
\end{equation}
with $j_l \in \{1,\ldots,M\}$, $s_l \in \{-1,1\}$, and  $q_l \in \{0,1\}$.
{The vectorial indices have been introduced to make the 
formulas more compact. The use of the indices is esemplificated by the 
following statement:
\begin{equation}
\begin{array}{l}
    \displaystyle\sum_{{\bf j}}\, (\cdot)=
    \displaystyle\sum_{j_{i_1}}\,
    \displaystyle\sum_{j_{i_2}}\,
    \ldots
    \displaystyle\sum_{j_{i_{N-1}}}\,
    \displaystyle\sum_{j_{i_{N}}}\,(\cdot).
\end{array}
\label{indecesVectorial}
\end{equation}
} 
Furthermore, we introduce two dimensionless variables, say $\xi$ and $\eta$,
in place of $p$ and $p'$, respectively, which are defined by
\begin{equation}
\begin{array}{l}
\xi=\sqrt 2\,\pi\,\sigma\,p,\\
\\
\eta=\sqrt 2\,\pi\,Q_I\,p',
\end{array}
\label{charFunction.8}
\end{equation}
so that Eq.~(\ref{charFunction.6}) becomes
\begin{equation}
\begin{array}{l}
\langle\exp({\rm i}2\pi Z)\rangle=\,\displaystyle\frac 1{2^M}\,
\exp\left[-(\xi^2+ \eta^2+2 \alpha\xi\eta)\right]\\
\\
\times
\displaystyle\sum_{{\bf j}}\,
\displaystyle\sum_{{\bf s}}\,
\displaystyle\sum_{{\bf q}}\,
\displaystyle\prod_{k\in {\cal N}}\,
(-1)^{q_k}\,\exp({\rm i}\beta_{j_k}\,s_k\,\phi_k\,\eta)\\
\\
\times
\,{\rm erf}\left(\hat a_{j_k+q_k}-{\rm i}\,s_k\,\phi_k\,\xi\right),
\end{array}
\label{charFunction.9}
\end{equation}
where 
\begin{equation}
\begin{array}{ccc}
\hat a_j=\displaystyle\frac{a_j}{\sigma\sqrt 2},&
\alpha=\displaystyle\frac{Q_I}{\sigma},&
\beta_j=y_j\,\displaystyle\frac{\sqrt 2 A_f}{Q_I}.
\end{array}
\label{charFunction.10}
\end{equation}

As far as the product $F_n(p)\,F_m(p')$ is concerned, from 
Eq. (\ref{FTquantization.1}) we have
\begin{equation}
\begin{array}{lr}
F_n(p)\,F_m(p')=
y_{M}^{n+m}\,\delta(\xi)\,\delta(\eta)\\
\\+
\displaystyle\sum_{j=1}^{M-1}\,y_{M}^{m}\,(y_{j}^{n}-y_{j+1}^{n})\,\displaystyle\frac{\sin(2\hat 
a_{j+1}\xi)}{\pi\xi}\,\delta(\eta)\\
\\+
\displaystyle\sum_{j=1}^{M-1}\,y_{M}^{n}\,(y_{j}^{m}-y_{j+1}^{m})\,\displaystyle\frac{\sin(2\hat 
a'_{j+1}\eta)}{\pi\eta}\,\delta(\xi)\\
\\+
\displaystyle\sum_{j=1}^{M-1}\,
\displaystyle\sum_{j'=1}^{M-1}\,
(y_{j}^{n}-y_{j+1}^{n})\,(y_{j'}^{m}-y_{j'+1}^{m})\,
\displaystyle\frac{\sin(2\hat 
a_{j+1}\xi)}{\pi\xi}\,\displaystyle\frac{\sin(2\hat a'_{j'+1}\eta)}{\pi\eta},
\end{array}
\label{charFunction.11a}
\end{equation}
for even values of both $n$ and $m$, and
\begin{equation}
\begin{array}{lr}
F_n(p)\,F_m(p')=
-y_{1}^{n+m}\,\displaystyle\frac{1}{\pi\xi}\,\displaystyle\frac{1}{\pi\eta}\\
\\+
\displaystyle\sum_{j=1}^{M-1}\,y_{1}^{m}\,(y_{j}^{n}-y_{j+1}^{n})\,
\displaystyle\frac{\cos(2\hat a_{j+1}\xi)}{\pi\xi}\,\displaystyle\frac{1}{\pi\eta}\\
\\+
\displaystyle\sum_{j=1}^{M-1}\,y_{1}^{n}\,(y_{j}^{m}-y_{j+1}^{m})\,
\displaystyle\frac{\cos(2\hat a'_{j+1}\eta)}{\pi\eta}\,\displaystyle\frac{1}{\pi\xi}\\
\\-
\displaystyle\sum_{j=1}^{M-1}\,
\displaystyle\sum_{j'=1}^{M-1}\,
(y_{j}^{n}-y_{j+1}^{n})\,(y_{j'}^{m}-y_{j'+1}^{m})\,
\displaystyle\frac{\cos(2\hat 
a_{j+1}\xi)}{\pi\xi}\,\displaystyle\frac{\cos(2\hat a'_{j'+1}\eta)}{\pi\eta},
\end{array}
\label{charFunction.11b}
\end{equation}
for odd values of both $n$ and $m$, where $\hat a_j'=\hat a/\alpha$.

Finally, on substituting from 
Eqs.~(\ref{charFunction.9}),~(\ref{charFunction.11a}) and~(\ref{charFunction.11b})
into Eq.~(\ref{ep.3}) we obtain Eqs.~(\ref{mixedMomentsOdd}) and~(\ref{mixedMomentsEven}), where
\begin{equation}
    \begin{array}{l}
    \mathcal{I}^{(e,1)}(\hat a)=\displaystyle\sum_{\mathbf{j},\mathbf{s,\mathbf{q}}}\,
    (-1)^{\mathbf{q}}\\
    \times\displaystyle\int\,
    \mathrm{d}\xi\,\exp(-\xi^{2})\,\displaystyle\frac{\sin(2\hat a\,\xi)}{\pi\xi}\,
    \displaystyle\prod_{k\in\mathcal{N}}\,\textrm{erf}(\hat 
    a_{j_{k}+q_{k}}-\mathrm{i}s_{k}\xi\phi_{k}),\\
    \\
    \\
    \mathcal{I}^{(e,2)}(\hat a')=\displaystyle\sum_{\mathbf{j},\mathbf{s,\mathbf{q}}}\,
    (-1)^{\mathbf{q}}\,
    \mathrm{erf}\left(\hat a'+\displaystyle\frac{\Gamma_{\bf s,j}}{2}\right)
    \displaystyle\prod_{k\in\mathcal{N}}\,\mathrm{erf}(\hat a_{j_{k}+q_{k}}),
    \\
    \\
    \mathcal{I}^{(e,3)}(\hat a,\hat a')=
    \displaystyle\sum_{\mathbf{j},\mathbf{s,\mathbf{q}}}\,
    (-1)^{\mathbf{q}}\,
    \mathrm{Re}
    \displaystyle\int\,
    \mathrm{d}\xi\,\exp(-\xi^{2})\,\displaystyle\frac{\sin(2\hat a\,\xi)}{\pi\xi}\\
    \times
    \mathrm{erf}\left(\hat a'+\displaystyle\frac{\Gamma_{\bf s,j}}{2}+\mathrm{i}\alpha\xi\right)
    \displaystyle\prod_{k\in\mathcal{N}}\,\textrm{erf}(\hat a_{j_{k}+q_{k}}-\mathrm{i}s_{k}\xi\phi_{k}),\\
    \\
    \\
    \mathcal{I}^{(o)}(\hat a,\hat a')=
    \displaystyle\sum_{\mathbf{j},\mathbf{s,\mathbf{q}}}\,
    (-1)^{\mathbf{q}}\,
    \mathrm{Im}
    \displaystyle\int\,
    \mathrm{d}\xi\,\exp(-\xi^{2})\,\displaystyle\frac{\cos(2\hat a\,\xi)}{\pi\xi}\\
    \times
    \mathrm{erf}\left(\hat a'+\displaystyle\frac{\Gamma_{\bf s,j}}{2}+\mathrm{i}\alpha\xi\right)
    \displaystyle\prod_{k\in\mathcal{N}}\,\textrm{erf}(\hat 
       a_{j_{k}+q_{k}}-\mathrm{i}s_{k}\xi\phi_{k}),\\
    \end{array}
    \label{charFunction.13}
\end{equation}
and
\begin{equation}
    \begin{array}{l}
    (-1)^{\mathbf{q}}=\displaystyle\prod_{k\in\mathcal{N}}\,(-1)^{q_{k}},\\
    \\
    \Gamma_{\bf s,j}=\displaystyle\sum_{k\in\mathcal{N}}\,s_{k}\beta_{j_{k}}\phi_{k}.
    \end{array}
    \label{charFunction.14}
\end{equation}

\section{Analysis of degradation effect in case a finite number of samples is used for signal reconstruction}
\label{finiteSamples}

The reconstruction formula in Eq.~(\ref{sampling.2.4.2.1}) 
assumes that an infinite number of samples can be used for estimating $w(\lambda)$. In practice, the sum will be made over a finite number of samples, applying some type of windowing 
function. For instance, in the case of a rectangular window one has simply that\footnote{In the present annex
we will indicate with tilde all terms which change due to the truncation.}
\begin{equation}
    \tilde w(\lambda) = A_f\,\displaystyle\sum_{i \in \mathcal{G}}\,
    u_i\,\phi_i,
    \label{finiteSamples.1}
\end{equation}
{The set $\mathcal{G}$ is made by consecutive indices distributed around 0, 
which we suppose including the set $\mathcal{N}$ defined in Eq. (\ref{setN}).}
In particular, it is straightforward to show that the constant $A_f$, 
which ensures minimum possible degradation, is still defined as in the 
case of infinite sum, i.e., by Eq.~(\ref{sampling.2.4.2.1.1}).

It is not difficult to show that the theoretical analysis developed 
still remains valid, provided that the random variable $Z_I$ defined 
in Eq.~(\ref{charFunction.2.1b}) be replaced by a new random variable, 
say $\tilde Z_I$, defined as the sum of two statistically independent 
terms, as follows:
\begin{equation}
\tilde Z_I=\displaystyle\sum_{{{k\notin \mathcal{N}}\atop{k\in \mathcal{G}}}} z_k+
\displaystyle\sum_{{k\notin \mathcal{G}}} p\,x_k\,\phi_k.
\label{finiteSamples.2}
\end{equation}
Then, on applying a similar methodology as done for the case of {$Z_I$, 
one can assume that $\langle\exp(\mathrm{i}2\pi\tilde Z_I)\rangle$ 
can be approximated by $\exp(-2\pi^2 \sigma_{\tilde Z_{I}}^2)$, where}
\begin{equation}
\sigma^2_{\tilde Z_I}=p^{2} P_{I} +  \tilde Q^2_{I}\,
(p^{\prime 2} + 2\,p\,p' ),
\label{finiteSamples.3}
\end{equation}
and
\begin{equation}
\tilde Q^2_{I}=
A_f\,\langle f^2\rangle\,
\left(
\displaystyle\sum_{k \in \mathcal{G}}\,\phi^2_k-
\displaystyle\sum_{k \in \mathcal{N}}\,\phi^2_k
\right),
\label{finiteSamples.4}
\end{equation}
provided that
\begin{enumerate}

\item a suitable set $\mathcal{N}$ has been selected according 
to the prescriptions of Sec.~\ref{results} (e.g. $\{0,1\}$);

\item $\mathcal{G}$ is much larger than $\mathcal{N}$.

\end{enumerate}
The whole theoretical analysis developed in the paper is now 
entirely applicable having care to replace $Q_I$ with $\tilde Q_I$.

To give a numerical evidence about the effect of the finite number 
of samples, Fig.~\ref{fig-fs.1} shows the behavior, as a function of 
the total number of samples, of the correlation coefficient, 
evaluated for $\lambda=1/2$ and for the 1-bit quantization function.
The dots are representative of the outcomes of numerical 
simulations,\footnote{Of course, the number of terms used for 
reconstructing $x(\lambda)$, according to Eq.~(2), was kept constant 
during all simulations to the relatively large of number of 500.} 
while the solid curve represents the results obtained by applying 
the theoretical analysis, together with the prescription given
by Eq.~(\ref{finiteSamples.4}). It is evident that the agreement 
between the values of $\rho$ obtained from  numerical simulations  
and those derived through the theoretical analysis in the present 
annex is quite satisfactory even when small number of terms is used 
for reconstruction. Additionally, it is also clear that selecting 
200 samples in the reconstruction formula is well representative of 
the ideal condition, since the values of $\rho$ have reached their 
asymptotic regime.

\newpage\

\newpage\

\section*{List of Tables}

\begin{table}[!ht]
    \begin{tabular}{l|l|l}
    \hline
    &&\\
    $M$ & $a_{j}\,\,(j=1,\ldots,M)$ & $y_{j}\,\,(j=1,\ldots,M)$ \\ 
    &&\\
    \hline  \hline
    &&\\
    1& $a_1=0.$ & $y_1=.798$\\
    &&\\
    \hline
    &&\\
    2& $a_1=0.$ & $y_1=.4528$\\
     & $a_2=0.9816$  & $y_2=1.510$\\
    &&\\
    \hline
    &&\\
    3& $a_1=0.$      & $y_1=0.3177$\\
     & $a_2=0.6589$  & $y_2=1.$\\
     & $a_3=1.447$  & $y_3=1.894$\\
    &&\\
    \hline
    &&\\
    4& $a_1=0.$      & $y_1=0.2451$\\
     & $a_2=0.5006$  & $y_2=0.7560$\\
     & $a_3=1.050$   & $y_3=1.344$\\
     & $a_4=1.748$   & $y_4=2.152$\\
    &&\\
    \hline
    &&\\
    \ldots & \ldots  & \ldots\\
    &&\\
    \hline
    \hline
    \end{tabular}
\caption{Quantization schemes, taken from Ref.~\cite{maxIREIT-60}, selected for testing of the theoretical approach. For meaning of symbols refer to Fig.~\ref{fig-levels}.}
\label{table.1}
\end{table}
\begin{table}[!ht]
    \begin{tabular}{|c|c|c|c|c|c|c|c|}
    \hline
    0.04
&   0
& 0 & 0 & 0 & 0 & 0 & 0\\
    \hline
    0
&   0.10
&   0.01
& 0 & 0 & 0 & 0 & 0\\
    \hline
    0 & 0.01
&   0.14
&   0.01
& 0 & 0 & 0 & 0\\
    \hline
    0 & 0 & 0.01
&   0.16
&   0.01
& 0 & 0 & 0\\
    \hline
    0 & 0 & 0 
&   0.01
&   0.16
&   0.01
& 0 & 0\\
    \hline
    0 & 0 & 0 & 0 
&   0.01
&   0.14
&   0.01
& 0 \\
    \hline
    0 & 0 & 0 & 0 & 0 
&   0.01
&   0.10
&   0
\\
    \hline
    0 & 0 & 0 & 0 & 0 & 0 
&   0
&   0.04
\\
    \hline
    \end{tabular}
\caption{Bivariate discrete probability distribution $p_{i,j}$,
defined as in Eq.~(\ref{bivariateDiscrete}), calculated for $M=4$ and
$\lambda=0.05$. indices $i$ (rows) and $j$ (columns) equal
$-M,\ldots,-1,1,\ldots,M$.}
\label{table.2a}
\end{table}
\begin{table}[!ht]
    \begin{tabular}{|c|c|c|c|c|c|c|c|}
    \hline
    0.03
&   0.01
&   0
&   0 & 0 & 0 & 0 & 0 \\
    \hline
    0.01 & 0.08 & 0.02 & 0 & 0 & 0 & 0 & 0 \\
    \hline
    0 & 0.02 & 0.12 & 0.03 & 0 & 0 & 0 & 0 \\
    \hline
    0 & 0 & 0.03 & 0.13 & 0.03 & 0 & 0 & 0 \\
    \hline
    0 & 0 & 0 & 0.03 & 0.13 & 0.03 & 0 & 0 \\
    \hline
    0 & 0 & 0 & 0 & 0.03 & 0.12 & 0.02 & 0 \\
    \hline
    0 & 0 & 0 & 0 & 0 & 0.02 & 0.08 & 0.01 \\
    \hline
    0 & 0 & 0 & 0 & 0 & 0 & 0.01 & 0.03\\
    \hline
    \end{tabular}
\caption{The same as in Table~\ref{table.2a} but for $\lambda=0.5$.}
\label{table.2b}
\end{table}
\begin{table}[!ht]
    \begin{tabular}{|c|c|c|c|c|c|c|c|}
    \hline
    0.04 & 0    & 0     & 0 & 0 & 0 & 0 & 0\\
    \hline
    0    & 0.10 & 0.01  & 0 & 0 & 0 & 0 & 0\\
    \hline
    0    & 0.01 & 0.14  & 0.01 & 0 & 0 & 0 & 0\\
    \hline
    0 & 0 & 0.01 & 0.16 & 0.01 & 0 & 0 & 0\\
    \hline
    0 & 0 & 0    & 0.02 & 0.16 & 0.01 & 0 & 0\\
    \hline
    0 & 0 & 0 & 0 & 0.01 & 0.14 & 0.01 & 0 \\
    \hline
    0 & 0 & 0 & 0 & 0 & 0.01 & 0.09 & 0 \\
    \hline
    0 & 0 & 0 & 0 & 0 & 0 & 0 & 0.04 \\
    \hline
    \end{tabular}
\caption{Bivariate discrete probability distribution $p_{i,j}$, 
estimated from numerical simulations for $M=4$ and $\lambda=0.05$,
to be compared to Tab.~\ref{table.2a}.}
\label{table.3a}
\end{table}
\begin{table}[!ht]
    \begin{tabular}{|c|c|c|c|c|c|c|c|}
    \hline
    0.03 & 0.01 & 0    & 0    & 0    & 0    & 0    & 0 \\
    \hline
    0.01 & 0.08 & 0.02 & 0    & 0    & 0    & 0    & 0 \\
    \hline
    0    & 0.02 & 0.12 & 0.02 & 0    & 0    & 0    & 0 \\
    \hline
    0    & 0    & 0.02 & 0.14 & 0.03 & 0    & 0    & 0 \\
    \hline
    0    &    0 & 0    & 0.03 & 0.14 & 0.02 & 0    & 0 \\
    \hline
    0    &    0 & 0    & 0    & 0.02 & 0.12 & 0.02 & 0 \\
    \hline
    0    &    0 & 0    & 0    & 0    & 0.02 & 0.08 & 0.01 \\
    \hline
    0    &    0 & 0    & 0    & 0    & 0    & 0.01 & 0.03\\
    \hline
    \end{tabular}
\caption{The same as in Table~\ref{table.3a} but for $\lambda=0.5$.
This table has to be compared to Tab.~\ref{table.2b}.}
\label{table.3b}
\end{table}

\newpage\
\newpage\

\section*{List of Figure Captions}

\begin{figure}[!ht]
\caption{Geometry of the quantization function.}
\label{fig-levels}
\end{figure}
\begin{figure}[!ht]
\caption{Theoretical behavior of the  cross-correlation coefficient, defined in Eq.~(\ref{rho}), 
as a function of $\lambda$, for $M=1,\ldots,4$.
}
\label{figCorrCoeff.1}
\end{figure}
\begin{figure}[!ht]
\caption{Block diagram showing the methodology adopted for
making the numerical simulations. The average in the $\langle \ldots \rangle$-block
was made over 10$^5$ realizations, while the sinc interpolation 
was performed by using 200 terms.
}
\label{fig-scheme-sim}
\end{figure}
\begin{figure}[!ht]
\caption{Comparison of the theoretical results shown in Fig.~\ref{figCorrCoeff.1}
(solid curves) to the experimental results obtained through numerical simulations (circles).}
\label{figSimulations.1}
\end{figure}
\begin{figure}[!ht]
\caption{Behavior of the correlation coefficient $\rho$ as a function of the total number of samples
used for reconstructing the values of $w(\lambda)$ for $\lambda=1/2$.  
Dots are representative of the numerical simulations; the solid curve represents
the results provided by the theoretical analysis with the use of Eq.~(\ref{finiteSamples.4}).
The quantization function corresponds to the 1-bit case.}
\label{fig-fs.1}
\end{figure}
\begin{figure}[!ht]
\caption{A DSP application. Sampling rate conversion applied to a signal $x(t)$ available through it quantized 
samples.}
\label{fig-src.1}
\end{figure}
\begin{figure}[!ht]
\caption{Behavior of the degradation between the sequences $\{u_m\}$ and $\{\tilde u_m\}$ in Fig.~\ref{fig-src.1},
experimentally evaluated for values of $L$  from 2 up to 30, and for several values of $D$.
The theoretical prediction given in Eq.~(\ref{src.4}) is also shown 
\textbf{(the theoretical points are joined by a solid curve which is used as a guide for the eye). The quantization functions correspond to $M=1$ (open circles),
 $M=2$ (black dots),  $M=3$ (open squares), and  $M=4$ (black squares).}}
\label{fig-src.2}
\end{figure}

\newpage\

\section*{Figures}

\centerline{\psfig{file=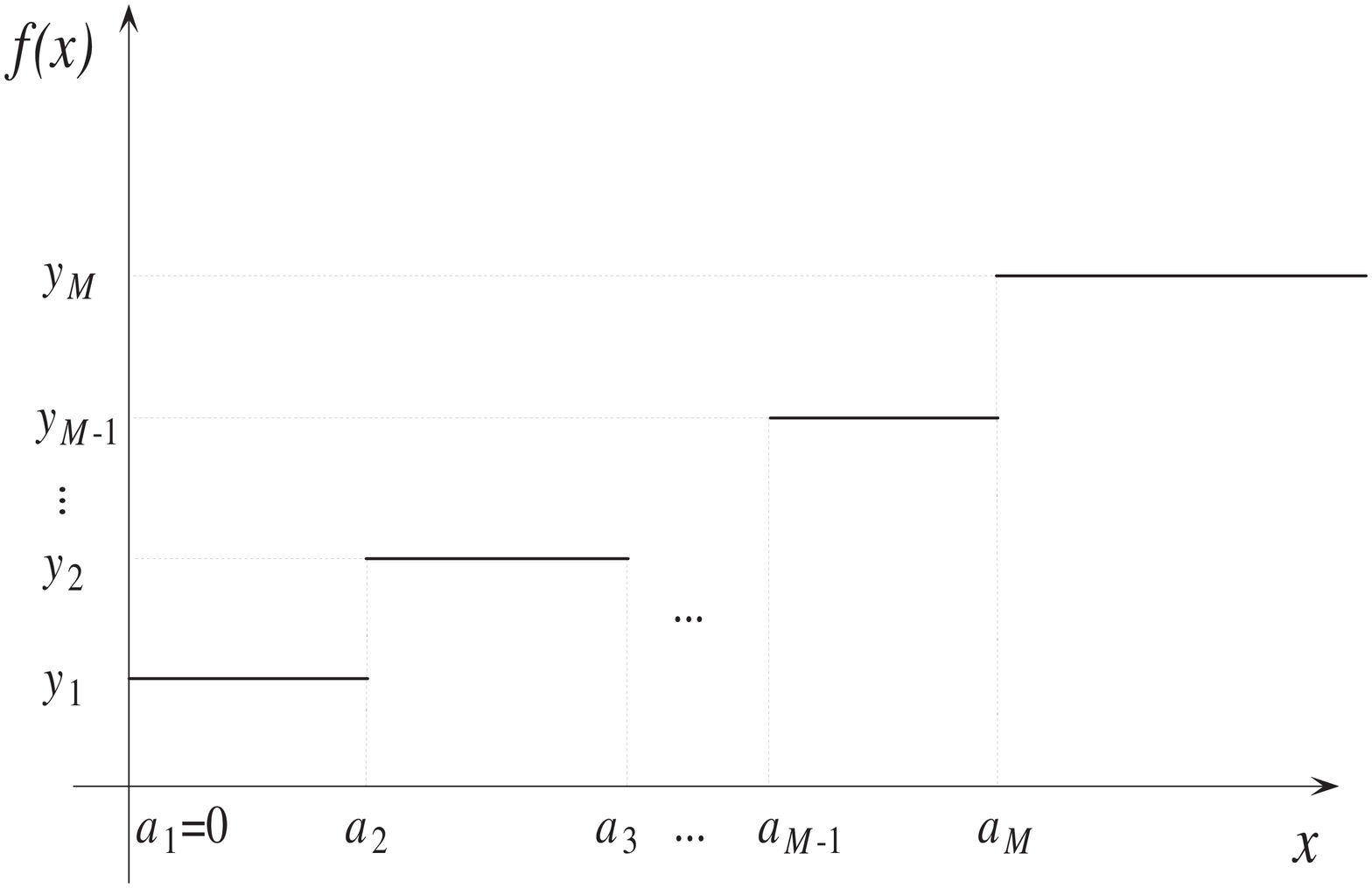,width=15cm,clip=,angle=0}}
\vspace*{1cm}
\centerline{Fig. 1 - Lanucara and Borghi}
\newpage\
%
\vspace*{1cm}
\centerline{Fig. 2 - Lanucara and Borghi}
\newpage\
\centerline{\psfig{file=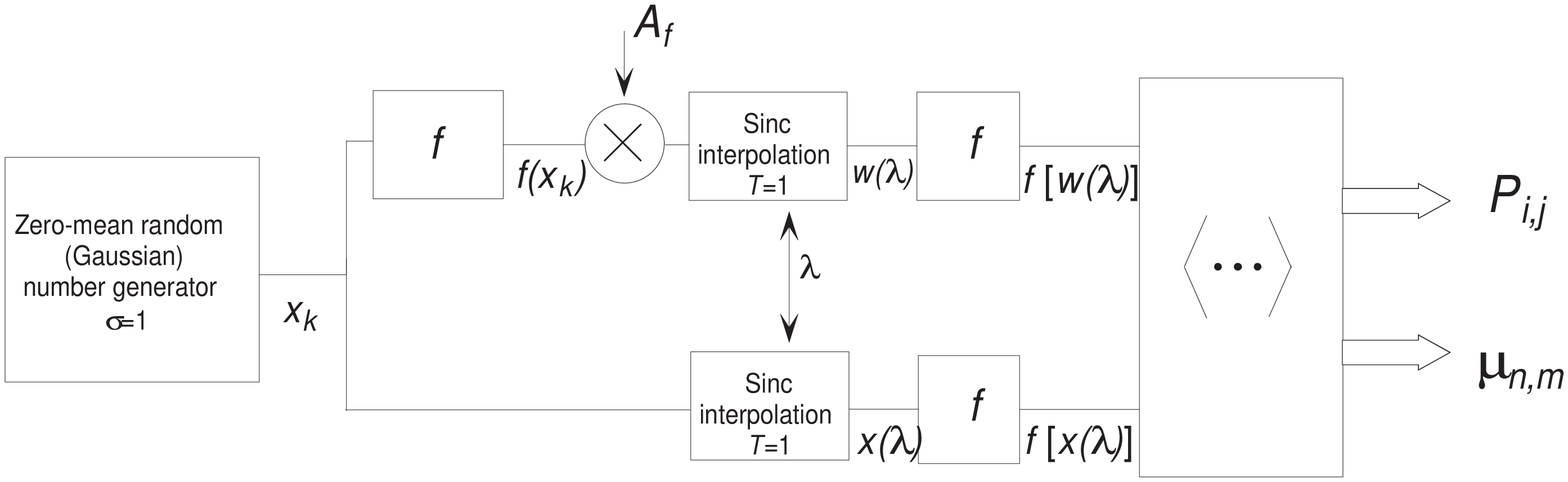,width=15cm,clip=,angle=0}}
\vspace*{1cm}
\centerline{Fig. 3 - Lanucara and Borghi}
\newpage\
\centerline{\psfig{file=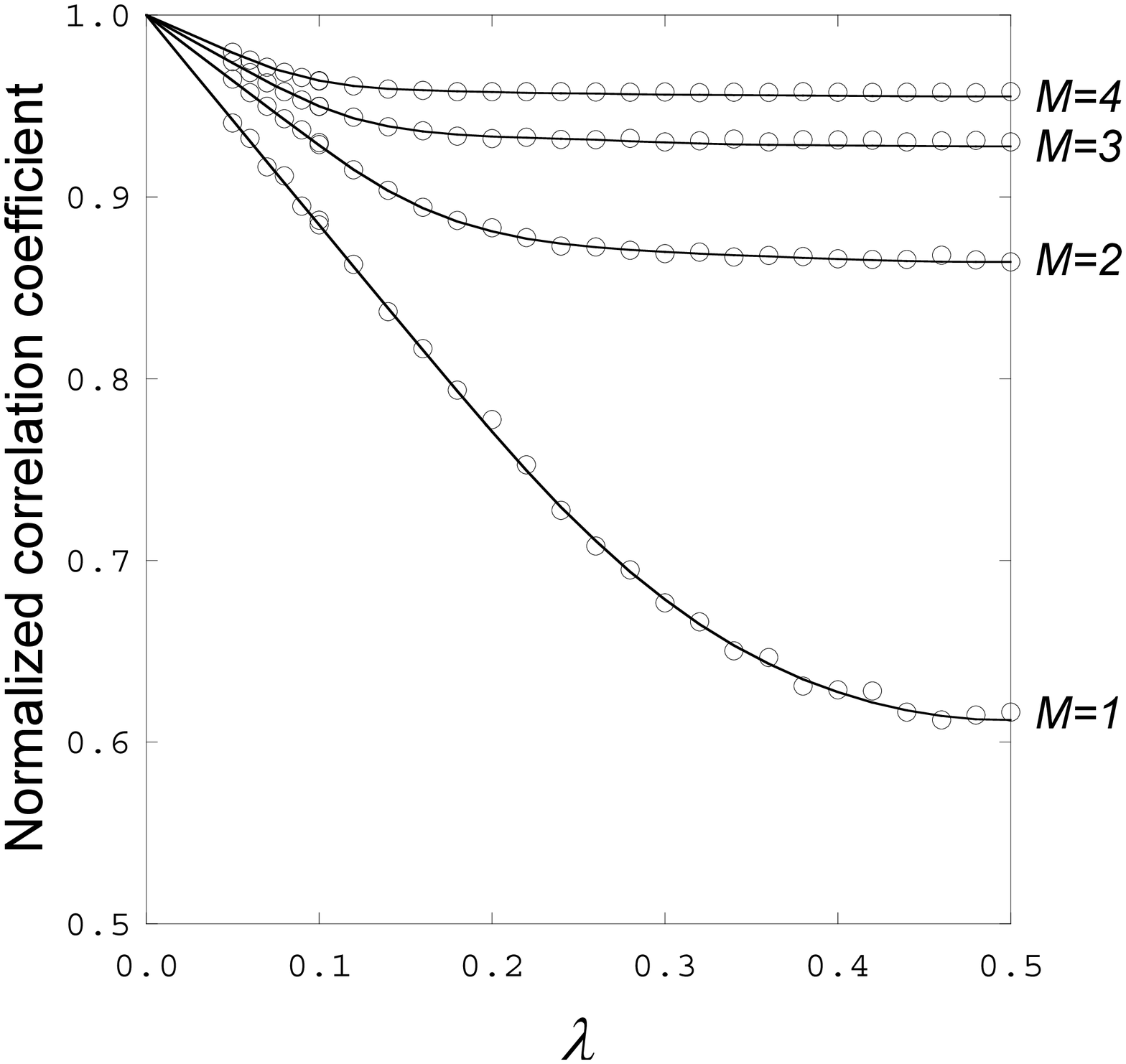,width=15cm,clip=,angle=0}}
\vspace*{1cm}
\centerline{Fig. 4 - Lanucara and Borghi}
\newpage\
\centerline{\psfig{file=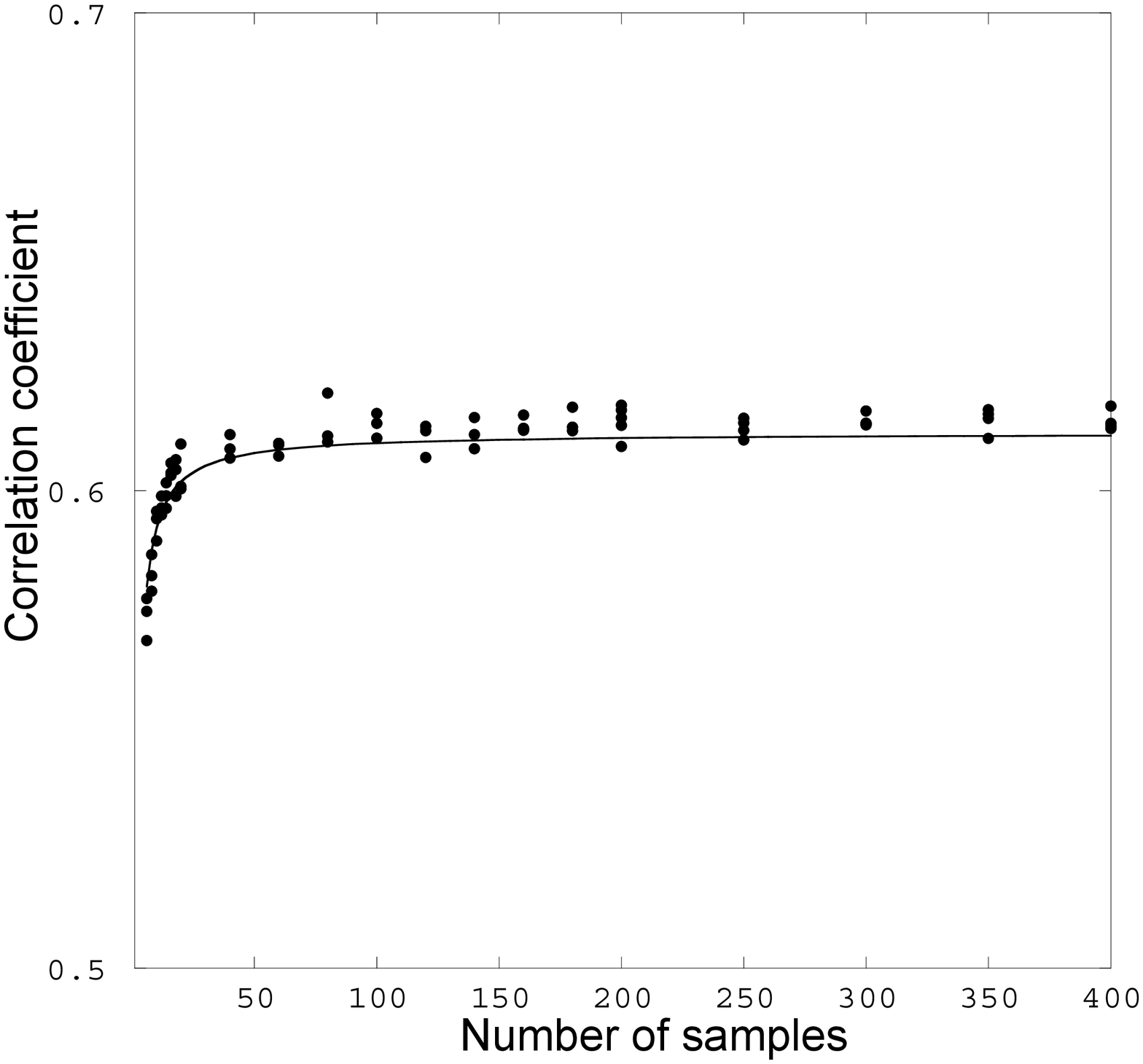,width=15cm,clip=,angle=0}}
\vspace*{1cm}
\centerline{Fig. 5 - Lanucara and Borghi}
\newpage\
\centerline{\psfig{file=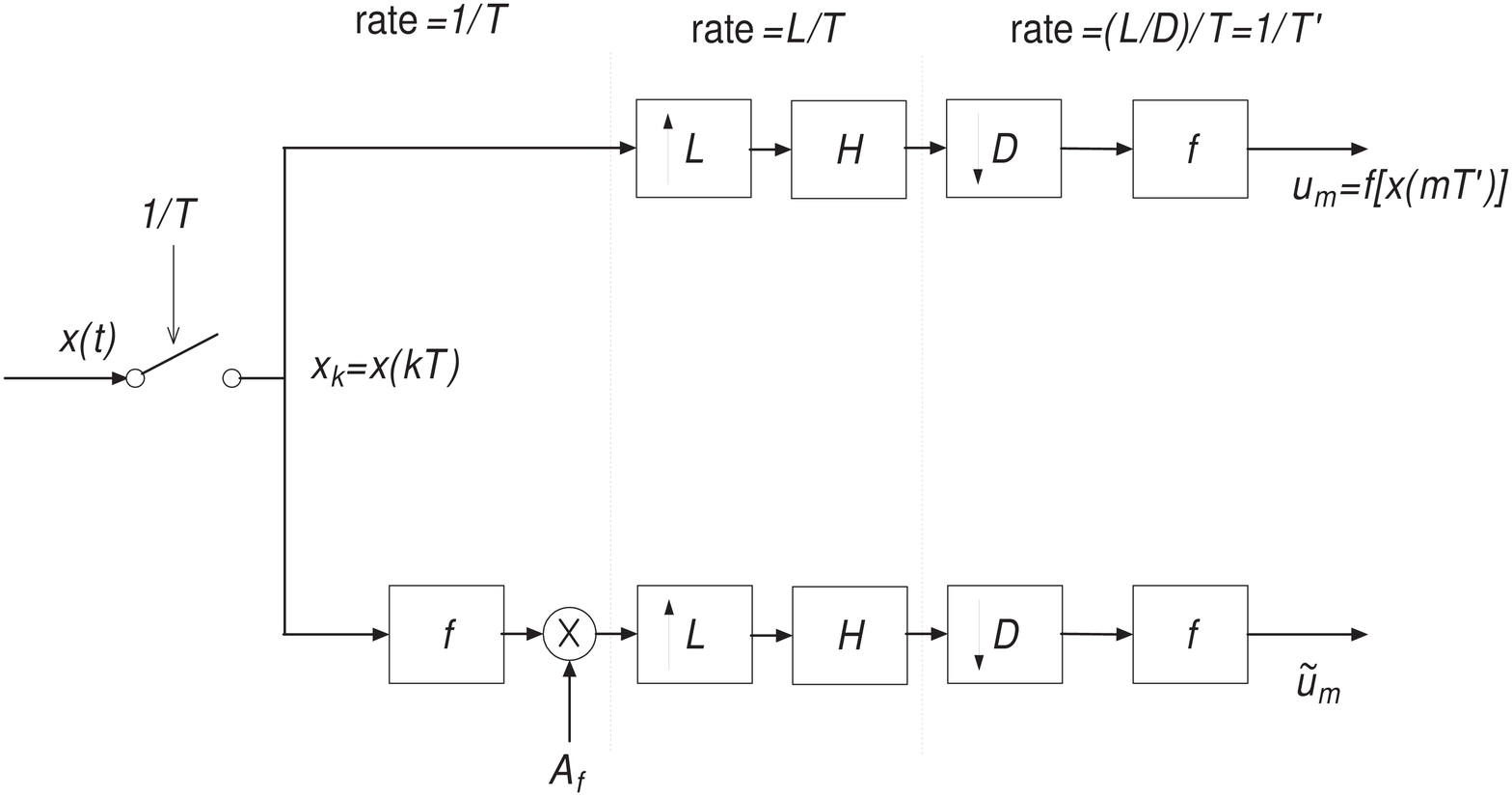,width=15cm,clip=,angle=0}}
\vspace*{1cm}
\centerline{Fig. 6 - Lanucara and Borghi}
\newpage\
\centerline{\psfig{file=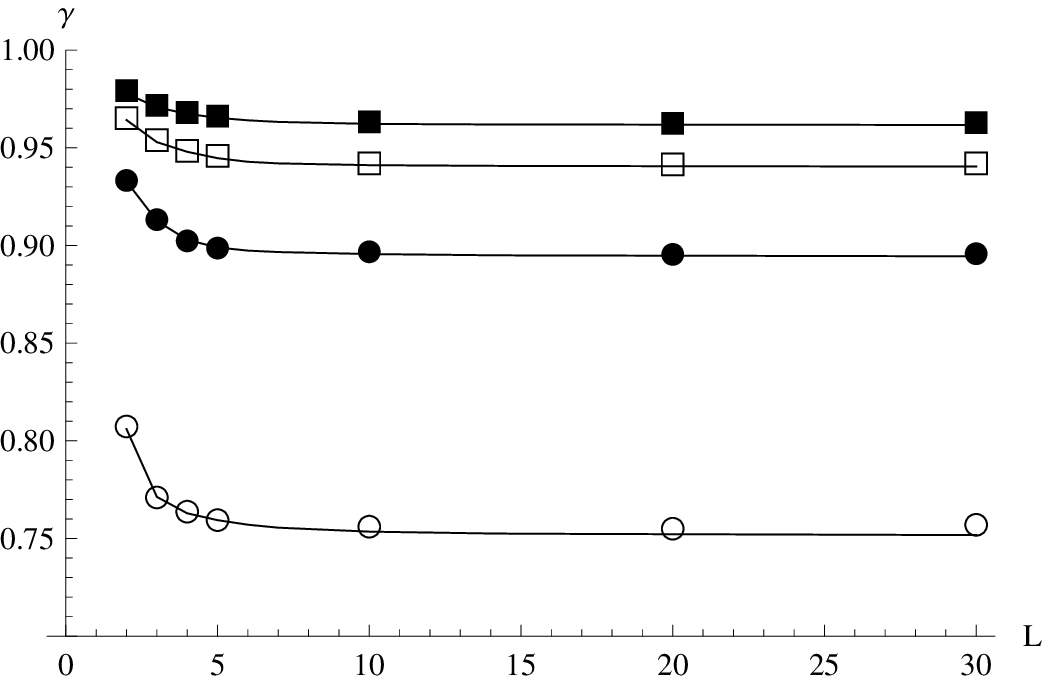,width=15cm,clip=,angle=0}}
\vspace*{1cm}
\centerline{Fig. 7 - Lanucara and Borghi}

\end{document}